\documentclass[lettersize,journal]{IEEEtran}
\usepackage{amsmath,amsfonts}
\usepackage{algorithmic}
\usepackage{algorithm}
\usepackage{array}
\usepackage[caption=false,font=normalsize,labelfont=sf,textfont=sf]{subfig}
\usepackage{textcomp}
\usepackage{stfloats}
\usepackage{url}
\usepackage{verbatim}
\usepackage{graphicx}
\usepackage{cite}
\usepackage{multirow}
\usepackage[table,xcdraw]{xcolor}
\usepackage[font=small,skip=0pt]{caption}
\usepackage{soul}
\usepackage{xcolor}
\usepackage{hyperref}

\begin{document}

\title{SCONNA: A Stochastic Computing Based Optical Accelerator for Ultra-Fast, Energy-Efficient Inference of Integer-Quantized CNNs}

\author{\IEEEauthorblockN{
Sairam Sri Vatsavai, Venkata Sai Praneeth Karempudi,   
Ishan Thakkar, Ahmad Salehi, and Todd Hastings}
\IEEEauthorblockA{\textit{Department of Electrical and Computer Engineering}} \\
University of Kentucky, Lexington, KY 40506, USA \\
\{ssr226, kvspraneeth, igthakkar, sayedsalehi, and todd.hastings\}@uky.edu}


\maketitle

\begin{abstract}
Convolutional Neural Networks (CNNs) are used extensively for artificial intelligence applications due to their record-breaking accuracy. For efficient and swift hardware-based acceleration, CNNs are typically quantized to have integer input/weight parameters. The acceleration of a CNN inference task uses convolution operations that are typically transformed into vector-dot-product (VDP) operations. Several photonic microring resonators (MRRs) based hardware architectures have been proposed to accelerate integer-quantized CNNs with remarkably higher throughput and energy efficiency compared to their electronic counterparts. However, the existing  photonic MRR-based analog accelerators exhibit a very strong trade-off between the achievable input/weight precision and VDP operation size, which severely restricts their achievable VDP operation size for the quantized input/weight precision of 4 bits and higher. The restricted VDP operation size ultimately suppresses computing throughput to severely diminish the achievable performance benefits. To address this shortcoming, we for the first time present a merger of stochastic computing and MRR-based CNN accelerators. To leverage the innate precision flexibility of stochastic computing, we invent an MRR-based optical stochastic multiplier (OSM). We employ multiple OSMs in a cascaded manner using dense wavelength division multiplexing, to forge a novel \underline{\textbf{S}}tochastic \underline{\textbf{C}}omputing based \underline{\textbf{O}}ptical \underline{\textbf{N}}eural \underline{\textbf{N}}etwork \underline{\textbf{A}}ccelerator (SCONNA). SCONNA achieves significantly high throughput and energy efficiency for accelerating inferences of high-precision quantized CNNs. Our evaluation for the inference of four modern CNNs at 8-bit input/weight precision indicates that SCONNA provides improvements of up to 66.5$\times$, 90$\times$, and 91$\times$  in frames-per-second (FPS), FPS/W and FPS/W/mm$^2$, respectively, on average over two photonic MRR-based analog CNN accelerators from prior work, with Top-1 accuracy drop of only up to 0.4\% for large CNNs and up to 1.5\% for small CNNs. We developed a transaction-level, event-driven python-based simulator for the evaluation of SCONNA and other accelerators (\url{https://github.com/uky-UCAT/SC_ONN_SIM.git}).
\end{abstract}

\section{Introduction}
Deep Neural Networks (DNNs) have revolutionized the implementation of various artificial intelligence tasks, such as image recognition, language translation, autonomous driving \cite{dnnapplications1,dnnapplications2}, due to their high inference accuracy. Convolutional Neural Networks (CNNs) are specific types of DNNs \cite{cnnapplication}. CNNs are computationally intensive, and hence, require a long inference time. In CNNs, around 80\% of the total processing time is taken by convolution operations that can be decomposed into vector dot product (VDP) operations \cite{Xu2018}. The ever-increasing complexity of CNNs has pushed for highly customized CNN hardware accelerators \cite{Baischer2021}. Often, for efficient and swift hardware-based acceleration, CNNs are typically quantized to have integer input/weight parameters \cite{krishnamoorthi2018quantizing}. Among CNN hardware accelerators, silicon-photonic accelerators have shown great promise to provide unparalleled parallelism, ultra-low latency, and high energy efficiency \cite{holylight,squeezelight,deapcnn,karen2020proceeding,amm,crosslight}. Typically, a silicon-photonic CNN accelerator consists of multiple Vector Dot Product Cores (VDPCs) that perform multiple VDP operations in parallel. Several VDPC-based optical CNN accelerators have been proposed in prior works based on various silicon-photonic devices, such as Mach Zehnder Interferometer (MZI) (e.g., \cite{mzi2018}, \cite{mzicomplex2021}, \cite{cansu2021}) and Microring Resonator (MRR) (e.g., \cite{deapcnn}, \cite{crosslight}, \cite{tait2014OSA}, \cite{tait2017}).

Among these optical VDPC-based CNN accelerators from prior work, the MRR-enabled VDPC-based accelerators (e.g., \cite{holylight,deapcnn,squeezelight,pixel,crosslight,tait2017}) have shown disruptive performance and energy efficiencies, due to the MRRs' compact footprint, low dynamic power consumption, and compatibility with cascaded dense-wavelength-division-multiplexing (DWDM). Among these MRR-enabled accelerators, some accelerators utilize digital VDPCs (e.g., \cite{pixel}), whereas some others employ analog VDPCs (e.g., \cite{crosslight,deapcnn,tait2017}). In general, a VDPC (analog or digital) transforms convolution operations into vector dot product (VDP) operations by decomposing the input tensors into vectors (1D tensors). In an analog VDPC, such VDP operations are also analog in nature, and they are performed on the individual VDP elements (VDPEs), which are the main MRR-enabled hardware components in the VDPCs. Multiple VDPEs in an analog VDPC can perform multiple analog VDP operations in parallel. The results of these analog VDP operations are converted into the digital format using analog-to-digital converters (ADCs). These results can be summed together (if and when needed) using a partial-sum (\textit{psum}) reduction network, which can be employed outside of the VDPCs as part of the post-processing components of the CNN accelerator. The functioning of the analog VDPCs and their constituent VDPEs in the ultra-high-speed, analog-optical domain results in disruptive throughput for performing analog VDP operations. 

We observe that two factors govern the performance of such analog optical VDPCs: (1) the achievable bit-precision (\textit{B}) and (2) the achievable scalability of the VDPCs, i.e., the achievable count of the individual VDPEs per VDPC (\textit{M}) and the individual VDPE size (the number of multiplications that can be generated and summed up per VDPE) (\textit{N}). In an analog VDPC, the achievable \textit{B} affects the inference accuracy of the processed CNNs, whereas the achievable VDPC scalability (i.e., \textit{N} and \textit{M}) directly defines the throughput of the VDPC for processing CNNs. Prior works \cite{lukasscalability} and \cite{Huang2020channelspacing} studied various factors such as optical power budget in waveguides, inter-channel spacing of wavelengths, crosstalk at cascaded MRRs, resolution of ADCs, and photodetector responsivity, to determine the bounds of the achievable B and scalability in analog optical VDPCs. Furthermore, prior work \cite{cases2022} characterized the very strong trade-off between the maximum achievable VDPC size \textit{N} and \textit{B} in analog optical VDPCs. From \cite{cases2022}, the analog optical VDPCs from prior works cannot support \textit{N} greater than 44 for B$>=$4-bit \cite{cases2022}. Achieving such low \textit{N} can seriously hurt the performance for processing modern CNNs. This is because modern CNNs employ tensors with as high as 4608 points (parameters) per tensor \cite{resnet}. Processing such large tensors on a VDPC with \textit{N}$\le$44 results in a large number of \textit{psums}, resulting in a very high latency overhead in the \textit{psum} reduction network. 
  
To avoid this undesired outcome, we advocate for such an architecture of MRRs-based CNN accelerator that achieves significantly larger VDPC size \textit{N} along with weakened interdependence between \textit{N} and \textit{B}. To that end, for the first time, we leveraged the inherent precision flexibility of stochastic computing to come up with a novel, MMRs-enabled \underline{\textbf{S}}tochastic \underline{\textbf{C}}omputing based \underline{\textbf{O}}ptical \underline{\textbf{N}}eural \underline{\textbf{N}}etwork  \underline{\textbf{A}}ccelerator (SCONNA). SCONNA employs our invented MRR-based Optical Stochastic Multipliers (OSMs) to realize manifold improvements in the throughput and energy efficiency of processing integer-quantized CNNs. 

Our key contributions in this paper are summarized below:

\begin{itemize}
    
     \item To enable stochastic computing in the optical domain, we present (i) a novel design of optical stochastic multiplier (OSM), and (ii) a novel photo-charge accumulator (PCA) circuit (Section \ref{section4});
     \item We present detailed modeling and characterization of our invented OSM and PCA using foundry-validated, commercial-grade, photonic-electronic design automation tools (Section \ref{section4}); 
    \item We employ our designed OSMs and PCAs to forge a highly scalable CNN accelerator named SCONNA, which employs OSM and PCA-based scalable VDPCs (Section \ref{section4});
    \item We perform a comprehensive scalability analysis for our SCONNA VDPCs, to determine their achievable maximum size \textbf{N}, operating speed, and error susceptibility (Section \ref{sectionV});
    \item We implement and evaluate SCONNA at the system-level using our in-house simulator (\url{https://github.com/uky-UCAT/SC_ONN_SIM.git}), and compare its performance and inference accuracy for processing 8-bit integer-quantized CNNs with two widely-known MRR-based analog CNN accelerators from prior works (Section \ref{evaluation}). 
\end{itemize}

\section{Preliminaries}
\subsection{Convolutional Neural Networks (CNNs)}
\begin{figure}[H]
  \centering
  \includegraphics[width=\linewidth]{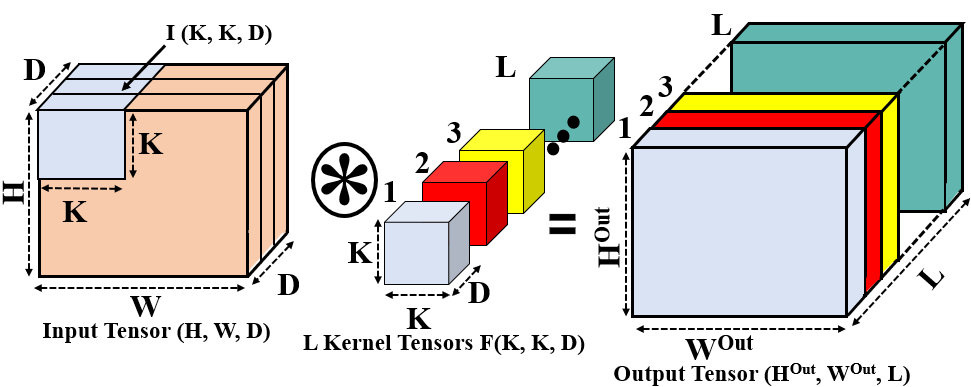}
  \caption{Illustration of a convolution operation.}
  \label{convolution}
\end{figure}

CNNs are specific types of DNNs that have shown remarkable accuracy for image classification. In general, a CNN consists of multiple convolutional layers, pooling layers, and fully connected layers. As shown in Fig. \ref{convolution}, a typical convolutional layer consists of one input tensor $\mathcal{I}$(\textit{H},\textit{W},\textit{D}) and  \textit{L} kernel tensors $\mathcal{F}$(\textit{K},\textit{K},\textit{D}). All of the \textit{L} kernel tensors convolve over the input tensor using stride ($\psi$) to produce the output tensor $\mathcal{O}$({\textit{H$^{Out}$},\textit{W$^{Out}$}},\textit{L}).

The computation required to produce each point \textit{O}(\textit{i}, \textit{j}, \textit{l}) in the output tensor $\mathcal{O}$(\textit{{H$^{out}$},\textit{W$^{out}$}},\textit{L}) can be given as Eq. \ref{convolution}. 

\begin{equation}
    O(i, j, l) = \sum_{d=1}^{D}\sum_{q=1}^{K}\sum_{r=1}^{K} F(r,q,d)I(i\times \psi+r,j\times \psi+q,d) 
    \label{convolutionequation}
\end{equation}
Here, \textit{d}=[1,\textit{D}], \textit{q}=[1,\textit{K}], \textit{r}=[1,\textit{K}], \textit{i}=[1,$H^{Out}$], \textit{l}=[1,\textit{L}], and \textit{j}=[1,$W^{Out}$] are various indices and their value ranges for the kernel and output tensors. \textit{O}(\textit{i},\textit{j},\textit{l}) in Eq. \ref{convolutionequation} is the sum of a total of \textit{K}$\times$\textit{K}$\times$\textit{D} products (products of the individual points of tensors $\mathcal{F}$ and $\mathcal{I}$(\textit{K},\textit{K},\textit{D}); $\mathcal{I}$(\textit{K},\textit{K},\textit{D}) is the gray-highlighted part of $\mathcal{I}$(\textit{H},\textit{W},\textit{D}) in Fig. \ref{convolution}). Thus, producing \textit{O}(\textit{i},\textit{j},\textit{l}) requires \textit{K}$\times$\textit{K}$\times$\textit{D} point-wise multiplications (to produce \textit{K}$\times$\textit{K}$\times$\textit{D} point-wise products) and one sum-of-products operation. The combination of these point-wise multiplications and the corresponding sum-of-products operation is mathematically equivalent to a Vector Dot Product (VDP) operation. A VDP operation typically occurs between two vectors. This implies that \textit{I} and \textit{F} in Eq. \ref{convolutionequation} are vectors, which are basically flattened (in 1D) versions of tensors $\mathcal{I}$(\textit{K},\textit{K},\textit{D}) and $\mathcal{F}$(\textit{K},\textit{K},\textit{D}) respectively. Note that vectors \textit{I} and \textit{K} have a total of \textit{S} = $\textit{K}\times\textit{K}\times\textit{D}$ points each. Henceforth, We refer to \textit{I} and \textit{K} as input vector and kernel vector, respectively.    

\subsection{Processing Convolutions on VDPCs}
Producing the output tensor $\mathcal{O}$(\textit{{H$^{Out}$,W$^{Out}$},L}) (Fig. \ref{convolution}) requires that the VDP operation shown in Eq. \ref{convolutionequation} is implemented multiple times, i.e., a total of $H^{Out} \times W^{Out} \times L$ times.
In Eq. \ref{convolutionequation}, the output \textit{O}(\textit{i},\textit{j},\textit{l}) is the result of the VDP operation between the corresponding input vector and kernel vector, each of size \textit{S} = $\textit{K}\times \textit{K}\times \textit{D}$ (Section II.A). Typically, for a CNN, the values \textit{K} and \textit{D} vary dramatically across different kernel tensors of the CNN. Therefore, \textit{S} = $\textit{K}\times\textit{K}\times\textit{D}$ also varies dramatically. The value \textit{S} for CNNs can be as large as 4608 (e.g., ResNet50 \cite{cases2022}). Because of such large \textit{S}, to accelerate VDP operations on a VDPC, it is intuitive to have the size \textit{N} of the constituent VDPEs of the VDPC (defined as the number of point-wise multiplications a VDPE can concurrently perform) to be as large as \textit{S}. However, it is hardly possible to have \textit{N} to be equal to \textit{S} in optical MRR-based analog VDPCs. Therefore, input vector and kernel vector are generally divided into multiple decomposed input vectors (\textit{DIV}s) (this and other abbreviations are defined in Table \ref{abbrevations}) and decomposed kernel vectors (\textit{DKV}s) first, and then these \textit{DIV}s and \textit{DKV}s are processed on the VDPEs (Section III.A). Having to decompose the input vector and kernel vector into multiple \textit{DIV}s and \textit{DKV}s raises several challenges as discussed in Section III.A.


\subsection{Optical Analog VDPC-Based CNN Accelerators}\label{analogaccelerators}\label{sec2c}
\begin{figure}[h!]
  \centering
  \includegraphics[width=\linewidth]{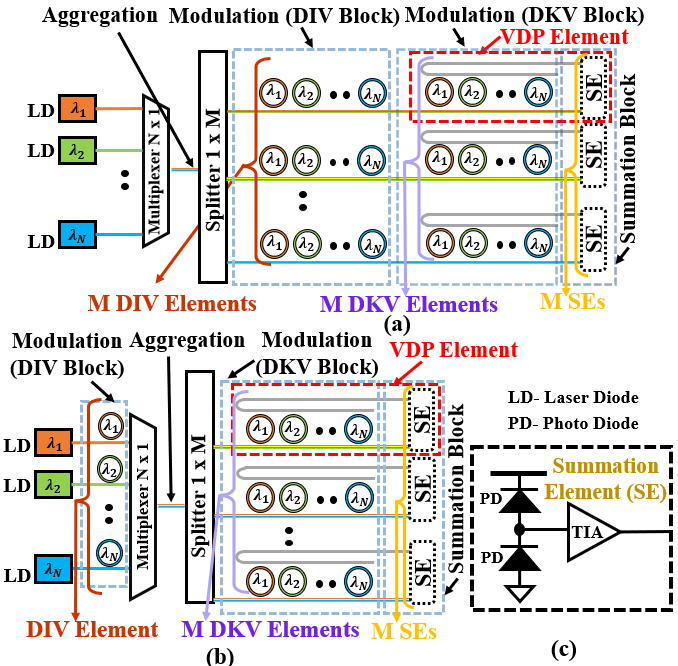}
  \caption{Illustration of common analog optical VDPC organizations: (a) AMM VDPC, (b) MAM VDPC. (c) Summation Element.}
  \label{AnalogTPCArchitecture}
\end{figure}

Most of the optical MRR-enabled analog, incoherent CNN accelerators from prior work employ multiple optical analog VDPCs that work in parallel. A brief review of prior works on optical accelerators is provided in Section \ref{relatedwork}. Typically, an analog VDPC implements the decomposed VDP operations of a convolution operation using DKVs and DIVs (Section II.A). In general, a VDPC consists of five blocks (Fig. \ref{AnalogTPCArchitecture}(a)): \textit{(i)} a laser block that consists of \textit{N} laser diodes (LDs) to generate \textit{N} optical wavelength channels; \textit{(ii)} an aggregation block that aggregates the generated optical wavelength channels into a single photonic waveguide through dense wavelength division multiplexing (DWDM) (using an \textit{N$\times$1} multiplexer) and then splits the optical power of these \textit{N} wavelength channels equally into \textit{M} separate waveguides (using a \textit{1$\times$M} splitter); \textit{(iii)} a modulation block, also referred to as \textit{DIV} block, that employs \textit{M} arrays of MRRs (one array per waveguide, with each array having \textit{N} MRRs; each array referred to as \textit{DIV} element) to imprint \textit{M} DIVs of \textit{N} points each onto the \textit{N$\times$M} wavelength channels by modulating the analog power amplitudes of the wavelength channels; \textit{(iv)} another modulation block, referred to as \textit{DKV} block, that employs another \textit{M} arrays of MRRs (one array per waveguide, with each array having \textit{N} MRRs; each array referred to as \textit{DKV} element) to further modulate the \textit{N$\times$M} wavelength channels with \textit{DKV}s, so that the analog power amplitudes of the individual wavelength channels then represent the point-wise products of the utilized \textit{DKV}s and \textit{DIV}s; and \textit{(v)} a summation block (SB) that employs a total of \textit{M} summation elements (SEs), with each SE having two balanced photodiodes (PDs) upon which the point-wise-product-modulated \textit{N} wavelength channels are incident to produce the output current that is proportional to the result of the VDP operation between the corresponding \textit{DKV} and \textit{DIV}. The laser block and SB are typically positioned at the two ends of the VDPC, with the aggregation, modulation (\textit{DIV}), and modulation (\textit{DKV}) blocks placed in between them.

Based on the order in which these intermediate blocks (aggregation, modulation (\textit{DIV}), modulation (\textit{DKV}) blocks) are positioned between the laser block and SB, we classify the MRR-based VDPC organizations from prior work as MAM (Modulation, Aggregation, Modulation) (e.g., \cite{holylight}, \cite{lukasscalability}) or AMM (Aggregation, Modulation, Modulation) (e.g., \cite{amm}, \cite{squeezelight}, \cite{deapcnn}). Fig. \ref{AnalogTPCArchitecture} illustrates MAM and AMM VDPC organizations. From Fig. \ref{AnalogTPCArchitecture}(a), the AMM VDPC organization positions the aggregation block first after the laser block, and then the \textit{DIV} modulation block followed by the \textit{DKV} modulation block. In contrast, the MAM VDPC in Fig. \ref{AnalogTPCArchitecture}(b) positions the \textit{DIV} modulation block first after the laser block, and then positions the aggregation block followed by the DKV modulation block. Note that the MAM \textit{DIV} block is structurally different from the AMM \textit{DIV} block. The MAM \textit{DIV} block employs only one MRR per waveguide, and as a result, it can imprint only one \textit{DIV} with \textit{N} points onto the \textit{N} wavelength channels. This one \textit{DIV} is shared among all DKVs in the MAM VDPC, whereas each \textit{DKV} can have a different \textit{DIV} corresponding to it in the AMM VDPC. Most MAM and AMM VDPCs from prior works have \textit{M=N}.

In both the AMM and MAM VDPC organizations, we refer to the combination of a \textit{DKV} element  and the corresponding SE as VDP element (VDPE). However, the size and point-wise product precision of MRR-based VDPEs have certain limitations (discussed in Section \ref{sec3}). These limitations demand exploration of new computing options to improve MRR-based VDPCs, and stochastic computing is an attractive option.

\subsection{Stochastic Computing}\label{stochasticprelim}
Stochastic Computing (SC) is an unconventional form of computing that represents and processes data in the form of probabilistic values called stochastic numbers (SNs) \cite{Gainesstochastic1969,alahistochasticsurvey2013,alaghistochastic2018}. In SC's unipolar format, an SN \textit{W} is a bit-stream of \textit{N} bits that represents a real-valued variable $\upsilon\in[0,1]$ by encoding $\upsilon$  through the ratio $N_1/N$, where $N_1$ is the number of 1's in \textit{W}. SC offers several advantages over conventional binary computing such as high error tolerance, low power consumption, small circuit area, and low-cost arithmetic operations consisting of standard digital logic components \cite{alaghistochastic2018}. For example, multiplication can be performed by a stochastic circuit consisting of a single AND gate. 

\begin{figure}[h!]
  \centering
  \includegraphics[width=\linewidth]{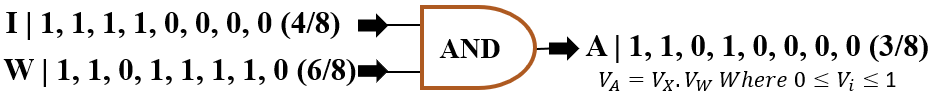}
  \caption{Multiplication between unipolar stochastic numbers \textit{I} and \textit{W}.}   
  \label{stochasticAND}  
\end{figure}
Fig. \ref{stochasticAND} illustrates a multiplication between two unipolar stochastic bit-streams \textit{I} and \textit{W} using an AND gate. The probabilities of seeing '1's in the bit-streams \textit{I} and \textit{W} are (4/8) and (6/8), respectively. The AND gate performs bit-wise logical AND operation on the bit-streams to produce the output bit-stream \textit{A}. In \textit{A}, the probability of seeing '1's is (3/8), which is equal to (4/8)$\times$(6/8), i.e., the multiplication (or product) of the input probabilities. Note that for the AND gate to produce an error-free multiplication output, the marginal probability of one bit-stream (i.e., \textit{I} or \textit{W}) should be equal to its conditional probability given the other bit-stream (i.e., \textit{I} given \textit{W} or \textit{W} given \textit{I})\cite{UGEMM}. Also note that, because of its advantages, SC has been adopted in stochastic deep CNNs \cite{stochasticCNNs,stochasticCNNs2,stochasticCNNs3}, GEMM computation \cite{UGEMM}, and image processing \cite{stochaticImageProcessing}. We use stochastic computing in this paper to relax the inherently strong scalability-precision trade-off in the optical VDPCs. This trade-off is explained in the next section.

\section{Motivation}\label{sec3}

\subsection{Scalability Limitations of MRR-Based Analog VDPCs}\label{sec3b}
 Prior works \cite{albireo}, \cite{cases2022}, and \cite{lukasscalability} have analysed the scalability (i.e., achievable value of VDPE size \textit{N} under the constraints of bit precision and data rate) of AMM and MAM VDPCs. Table \ref{scalability} reproduces the supported values of VDPE size \textit{N} (considering \textit{M=N}) for AMM and MAM VDPCs at various data rates (DRs) and bit precision from \cite{cases2022}. From Table 1, the maximum \textit{N}=44 is obtained for MAM VDPC across all tested DR and B values. For MAM VDPC for 1 GS/s, maximum \textit{N} reduces from 44 to 12 as we increase the input/weight precision from 4-bit to 6-bit. The reason for such strong trade-off between \textit{N} and achievable input/weight precision (referred to as \textit{B}, henceforth) in MAM and AMM VDPCs is that both \textit{B} and \textit{N} strongly depend on the number of distinguishable analog optical power levels \cite{cases2022}\cite{albireo}, which is proportional to \textit{$N\times2^B$}. Hence, for a fixed number of distinguishable analog optical power levels, which is defined by the analog optical power resolution of the utilized summation elements (SEs) (see SEs in Fig. \ref{AnalogTPCArchitecture}) the supported \textit{N} drastically decreases with an increase in \textit{B}. As a result, \textit{N} decreases all the way to 1 when \textit{B} increases to 8-bit \cite{cases2022}. 
 
 Due to such strong trade-off between \textit{N} and \textit{B}, the MAM and AMM type of analog VDPCs face two consequences. First, they produce high number of partial sums and incur significantly high latency for partial sum reduction. For example, a VDPE with \textit{N}=44 for \textit{B}=4-bit can only produce a VDP operation between two 44-point vectors. 
 Therefore, producing a VDP operation between an input vector and kernel vector with \textit{S}=4608 (e.g., ResNet50 \cite{resnet}\cite{cases2022}) requires that the input vector is first decomposed into a total of \textit{C}=Ceil(S/N)=105 \textit{DIV}s of \textit{N}=44 points each. Similarly, the kernel vector also needs to be decomposed into a total of \textit{C}=Ceil(S/N)=105 \textit{DKV}s of \textit{N}=44 points each. Then, a total of 105 VDPEs can be employed to perform 105 VDP operations between 105 pairs of \textit{DKV}s and \textit{DIV}s, to consequently produce a total of 105 intermediate VDP results (i.e., partial sums (\textit{psums})). Although these 105 VDP operations can be parallelized over 105 VDPEs, producing the final VDP result of \textit{S}=4608 would require the accumulation of the 105 \textit{psums}. Doing so can incur very high latency and energy consumption, which should be avoided using a more efficient VDPC design. 
 
 \begin{table}[]
\centering
\caption{VDPE size \textit{N} for input/weight precision=\{4,6\}-bit at data rates (DRs)=\{1,3,5,10\}GS/s, for AMM and MAM VDPCs.}
\label{scalability}
\begin{tabular}{|c|c|cccc|}
\hline
\multirow{2}{*}{\textbf{VDPC}}                      & \multirow{2}{*}{\textbf{Precision}} & \multicolumn{4}{c|}{\textbf{Datarate(DR)}}                                                                                                \\ \cline{3-6} 
                                                    &                                         & \multicolumn{1}{c|}{\textbf{1 GS/s}} & \multicolumn{1}{c|}{\textbf{3 GS/s}} & \multicolumn{1}{c|}{\textbf{5 GS/s}} & \textbf{10 GS/s} \\ \hline
\multirow{2}{*}{\textbf{AMM}}                       & \textbf{4-bit}                          & \multicolumn{1}{c|}{31}              & \multicolumn{1}{c|}{20}              & \multicolumn{1}{c|}{16}              & 11               \\ \cline{2-6} 
                                                    & \textbf{6-bit}                          & \multicolumn{1}{c|}{6}               & \multicolumn{1}{c|}{3}               & \multicolumn{1}{c|}{2}               & 1                \\ \hline
\multicolumn{1}{|l|}{\multirow{2}{*}{\textbf{MAM}}} & \textbf{4-bit}                          & \multicolumn{1}{c|}{44}              & \multicolumn{1}{c|}{29}              & \multicolumn{1}{c|}{22}              & 16               \\ \cline{2-6} 
\multicolumn{1}{|l|}{}                              & \textbf{6-bit}                          & \multicolumn{1}{c|}{12}              & \multicolumn{1}{c|}{7}               & \multicolumn{1}{c|}{5}               & 3                \\ \hline
\end{tabular}
\end{table}

\begin{table}[]
\centering
\caption{Total number of kernels (\textit{$T_L$}) of different DKV sizes (\textit{S}) for various CNNs. The \textit{$T_L$} values were extracted for trained CNN models from Keras Applications \cite{chollet2015keras}.}
\label{CNNTensorInfo}
\begin{tabular}{|c|c|c|c|c|c|}
\hline
\textbf{Model}                     & \textbf{$T_L$} & \textbf{S}        & \textbf{Model}                      & \textbf{$T_L$} & \textbf{S}        \\ \hline
\multirow{2}{*}{\textbf{ResNet50}} & 1          & S$\leq$44    & \multirow{2}{*}{\textbf{GoogleNet}} & 13         & S$\leq$44    \\ \cline{2-3} \cline{5-6} 
                                   & 26562      & S\textgreater{}44 &                                     & 7554       & S\textgreater{}44 \\ \hline
\multirow{2}{*}{\textbf{VGG16}}    & 69         & S$\leq$44    & \multirow{2}{*}{\textbf{DenseNet}}  & 1          & S$\leq$44    \\ \cline{2-3} \cline{5-6} 
                                   & 4168       & S\textgreater{}44 &                                     & 10242      & S\textgreater{}44 \\ \hline
\end{tabular}
\end{table}
 
As the second consequence, the throughput of the MAM and AMM VDPCs decreases at higher bit precision (higher value of \textit{B}). This is because to avoid a drastic decrease in \textit{N} as \textit{B} increases beyond 4-bit, the AMM and MAM type of analog VDPCs typically operate at \textit{B}=4-bit \cite{cases2022}. However, using at least 8-bit precision (\textit{B}=8-bit) for the integer-quantized CNN models is recommended to achieve competitive inference accuracy, while also reducing the computational effort, memory requirements, and energy consumption \cite{krishnamoorthi2018quantizing}. To meet this requirement, analog VDPCs from \cite{holylight} (an MAM VDPC) and \cite{deapcnn} (an AMM VDPC) employ bit-slicing of input/weight parameters. They slice each 8-bit integer input/weigh parameter into two slices of 4-bit each. Then, they employ two VDPCs in parallel; each VDPC processes one 4-bit slice of the input/weight parameters. The corresponding 4-bit VDP results from these two 4-bit VDPCs are then combined using shifters and adders to produce the final 8-bit results. Thus, performing VDP operations using bit slices reduces the total number of VDP results that can be produced by a fixed number of VDPCs, because multiple VDPCs are needed to produce a single set of VDP results. This can severely degrade the throughput of such VDPCs. Such undesired outcome should be avoided by designing a more efficient VDPC.

\subsection{Need for Stochastic Computing}
Table \ref{CNNTensorInfo} reports the counts of kernel tensors according to their sizes \textit{S} (\textit{S}$<=$44 and \textit{S}$>$44) for four modern CNN models. From Table \ref{CNNTensorInfo}, more than 98\% of the kernel tensors across all four CNNs have \textit{S}$>$44, and thus, they require VDPEs with size \textit{N}$>$44 to process their corresponding VDP operations. But, from Table \ref{scalability}, the maximum achievable \textit{N} for analog VDPCs at 4-bit precision (\textit{B}=4-bit) is limited to 44; therefore, processing the VDP operations corresponding to more than 98\% of kernel tensors that have \textit{S}$>$44 would lead to high \textit{psum} reduction latency (see Section \ref{sec3b}). However, reducing this \textit{psum} reduction latency in analog VDPCs is challenging, as they have a strong trade-off between \textit{N} and \textit{B}, and this is because the required number of analog optical power levels (i.e., $2^B$) to support a specific \textit{B} consumes a large part of the available dynamic range of optical power in analog VDPCs. 
To this end, the remaining dynamic range of optical power within the total allowable optical power budget restricts the supported \textit{N} in analog VDPCs. This limitation can be addressed by performing VDP operations in the digital domain \cite{pixel}. There is no need to support any analog optical power levels in the digital domain; therefore, most of the available dynamic range of optical power in a digital VDPC can be used to support a higher \textit{N}. But, the MRR-based binary digital VDPCs (e.g., \cite{pixel,shreyan2021}) suffer from very high hardware complexity, and their multiply-accumulate and bit-shifting circuits consume huge area. These drawbacks motivate the need to examine new options for realizing optical digital VDPCs.

One such option is stochastic computing. In stochastic computing, multiplications can be replaced with simple bitwise AND operations \cite{alaghistochastic2018}. This can be leveraged to perform point-wise multiplications between \textit{DKV}s and \textit{DIV}s (Section \ref{sec2c}) with less hardware complexity than binary digital VDPCs. In addition, since stochastic computing is also implemented in the digital domain (non-binary), a stochastic computing based optical VDPC can support a large \textit{N} due to the large available dynamic range of optical power, just as discussed above for a binary digital VDPC. Moreover, a stochastic computing based optical VDPC can attain different precision levels
by merely changing the number of bits in the stochastic bit-streams, without requiring different analog optical power levels. Therefore, to utilize these advantages of stochastic computing, prior works \cite{ZhangISCA2018} and \cite{ElDerhalli2021} proposed stochastic computing based
photonic acceleration. \cite{ZhangISCA2018} reports acceleration of Markov Random Field Inference and \cite{ElDerhalli2021} employs photonic crystals and MZIs to build an edge detection filter. However, none of the prior works have employed stochastic computing based photonic acceleration for neural network inference. To fill this gap, we invent an MRR-based optical stochastic multiplier (OSM) and employ multiple OSMs to forge a novel Stochastic Computing Optical Neural Network Accelerator (SCONNA). The following section discusses our SCONNA architecture.

\section{Our Proposed SCONNA Architecture}\label{section4}
\subsection{Overview of SCONNA VDPC}\label{sconna-vdpc-overview}
\begin{figure*}[h!]
  \centering
  \includegraphics[scale=0.43]{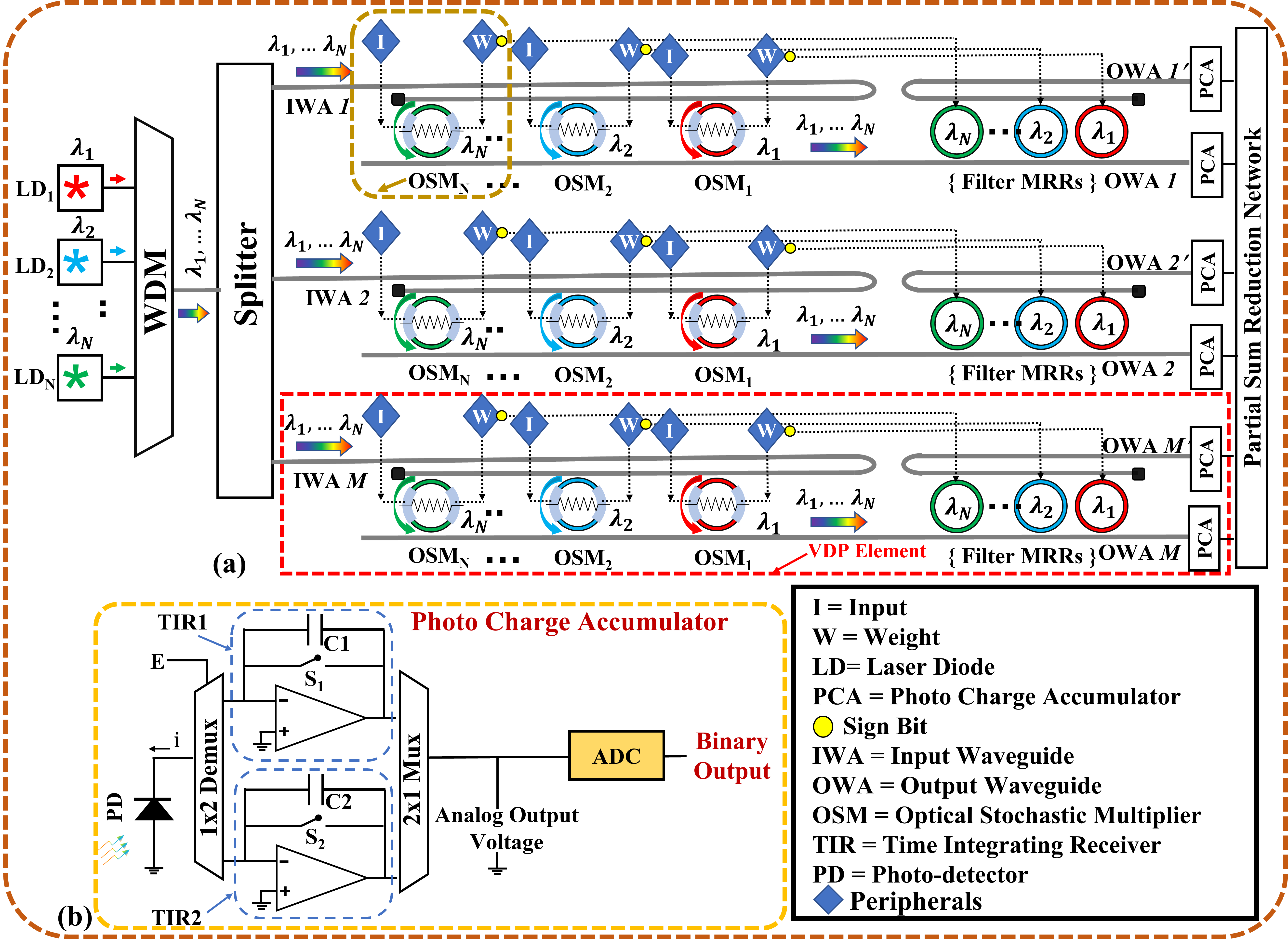}
  \caption{Schematics of (a) Our SCONNA VDPC  (b) Photo-Charge Accumulator (PCA) Circuit.}  
 \label{SCONNA_VDPC}
\end{figure*}

Fig. \ref{SCONNA_VDPC}(a) illustrates the VDPC organization of our SCONNA architecture. Like the VDPCs of analog optical accelerators, a SCONNA VDPC also implements multiple VDP operations in parallel. For that, an array of total \textit{N} single-wavelength laser diodes (LDs) are used, with each LD sourcing optical power of $P_{\lambda_i}^{in}$ amount at a distinct wavelength $\lambda_i$. The total power from all \textit{N} LDs ($\lambda_1$ to $\lambda_N$) multiplexed into a single photonic waveguide through wavelength division multiplexing (WDM). These multiplexed wavelengths split into \textit{M} input waveguide arms (IWAs). Every IWA receives \textit{N}-wavelength optical power and guides it to a VDPE. There are a total of \textit{M} IWAs and \textit{M} VDPEs in the SCONNA VDPC (Fig. \ref{SCONNA_VDPC}(a)).

Each VDPE consists of three components: (i) a cascade of \textit{N} Optical Stochastic Multipliers (OSMs); (ii) a bank of filter MRRs; (iii) a Photo-Charge Accumulator (PCA) pair. Each OSM performs stochastic multiplication between an input bit-stream \textit{I} (corresponding to a point in an \textit{N}-point DIV) and weight bit-stream \textit{W} (corresponding to a point in an \textit{N}-point DKV).
Each OSM receives its bit-streams \textit{I} and \textit{W} from its corresponding peripherals at a supported bitrate (\textit{BR}). Bit-stream \textit{W} provides a weight value along with a sign bit. Bit-stream \textit{I} provides the RELU-activated output value from the previous CNN layer, without a sign bit as RELU has a non-negative output. The detailed design of \textit{OSM}s and their peripherals is explained in Section \ref{osmgate}. Each OSM performs a bit-wise logical AND operation between the \textit{I} and \textit{W} bit-streams to produce a resultant optical bit-stream that represents the stochastic multiplication between the \textit{I} and \textit{W} bit-streams. The \textit{N} optical bit-streams from the cascade of \textit{N} OSMs, with each bit-stream carrying a stochastic multiplication result, reach the bank of filter MRRs. In this bank, each filter MRR operates on a distinct optical bit-stream $\lambda_i$. Each filter MRR receives the sign bit from the peripheral \textit{W} of its corresponding OSM (Fig. \ref{SCONNA_VDPC}(a)). The sign bit operates the filter to steer the incoming optical bit-stream $\lambda_i$ to the output waveguide arm \textit{OWA} (if the sign bit is '0') or \textit{OWA'} (if the sign bit is '1'). Thus, the OWA and OWA' of a VDPE guide the optical bit-streams, carrying the stochastic multiplication results, to PCAs. A PCA is a circuit that collects all the optical bit-streams (i.e., stochastic multiplication results) from its corresponding OWA (or OWA') and generates the accumulation result in the binary format (details about PCA in Section \ref{pcacircuit}). In a VDPE, the OWA-coupled PCA combines with the corresponding OWA'-coupled PCA to generate a signed accumulation result.    

\subsection{Optical Stochastic Multiplier}\label{osmgate}
\begin{figure}[H]
  \centering
    \includegraphics[scale=0.55]{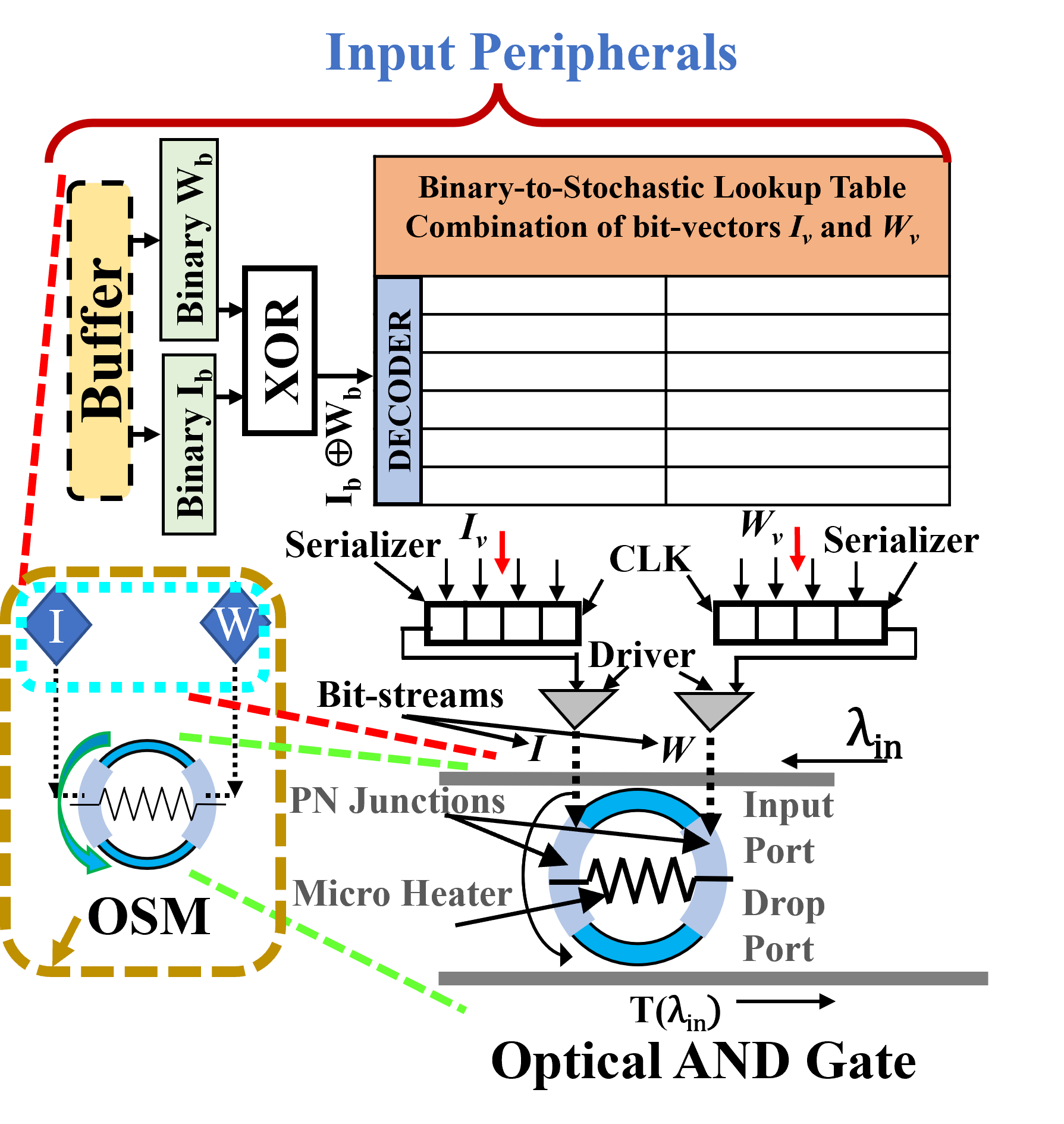}
  \caption{Schematic of our Optical Stochastic Multiplier (OSM). }   
\label{OSM} 
\end{figure}

Our Optical Stochastic Multiplier (OSM) consists of peripherals and an Optical 'AND' Gate (OAG) (Fig. \ref{OSM}). The peripherals convert a binary input value $I_b$ and binary weight value $W_b$ into unipolar stochastic bit-streams \textit{I} and \textit{W}, and OAG performs multiplication-equivalent bitwise AND operation between the stochastic bit-streams \textit{I} and \textit{W}.

From Fig. \ref{OSM}, the peripherals of our OSM use a lookup table and serializers to generate a combination of unipolar stochastic bit-streams \textit{I} and \textit{W}. From \cite{UGEMM}, two unipolar stochastic bit-streams, for their eventual error-free multiplication using an AND gate, should be generated in combination with each other, so that they are uncorrelated, i.e., the marginal probability of one bit-stream (i.e., \textit{I} or \textit{W}) is equal to its conditional probability given the other bit-stream (i.e., \textit{I} given \textit{W} or \textit{W} given \textit{I}). For our OSM, we propose to generate all possible combinations of uncorrelated bit-streams \textit{I} and \textit{W} a priori (offline) using the unipolar circuit from \cite{UGEMM}, and then store these bit-streams in the bit-vector (bit-parallel) format in the lookup table (Fig. \ref{OSM}). As a result, each entry in the lookup table stores a combination of uncorrelated bit-vectors $I_v$ and $W_v$. To index into this lookup table, our OSM creates a unique identifier for each combination of binary values $I_b$ and $W_b$ (that are accessed from a buffer (a scratchpad memory); Fig. \ref{OSM}) by performing an XOR-based hash function $I_b \oplus W_b$. Thus, our OSM uses a $I_b \oplus W_b$ value to fetch the desired combination of $I_v$ and $W_v$ from the lookup table. Then, it pushes these $I_v$ and $W_v$ through dedicated high-speed serializers, to generate bit-streams \textit{I} and \textit{W}. \underline{Lookup table size}: If precision \textit{B}=8-bit for binary $I_b$ and $W_b$, there are $2^B$ entries in the lookup table, with each entry storing two $2^B$-bits long bit-vectors.

\begin{figure}[h!]
  \centering
  \includegraphics[width=\linewidth]{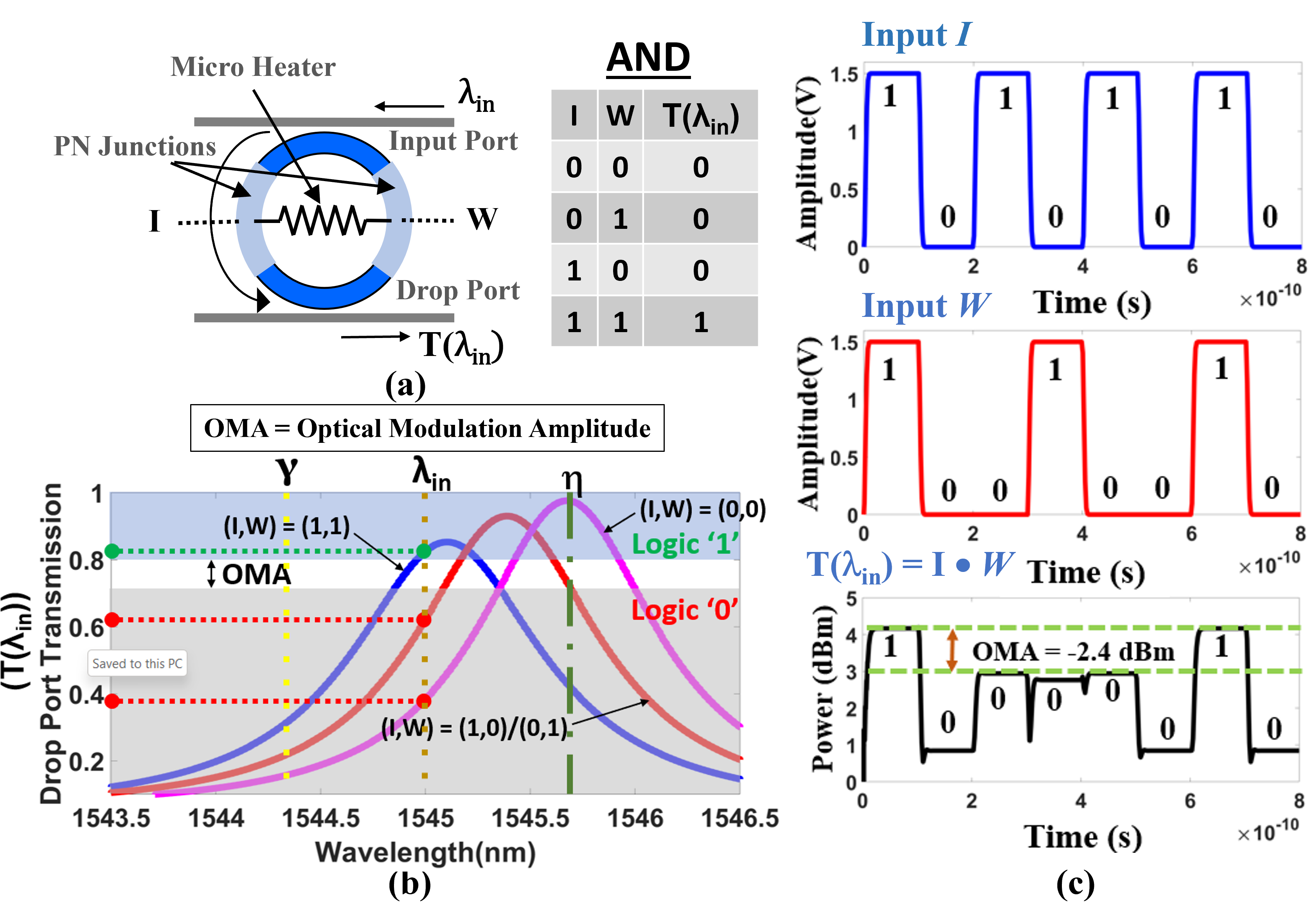}
  \caption{(a) Schematic of our Optical AND Gate (OAG), (b) operation of OAG, (c) results of OAG's transient analysis.}   
\label{OSMG_gate} 
\end{figure}

The stochastic bit-streams \textit{I} and \textit{W}, generated by the peripherals of our OSM, are then fed to the OAG via high-speed drivers for their stochastic multiplication (Fig. \ref{OSM}). The design of our OAG is illustrated in Fig. \ref{OSMG_gate}(a). Our OAG is an add-drop microring resonator (MRR), which has two operand terminals (realized as embedded PN-junctions) that can take two stochastic bit-streams \textit{I} and \textit{W} (Fig. \ref{OSMG_gate}(a)) as inputs at a predefined bitrate (BR). Fig. \ref{OSMG_gate}(b) shows the passbands of the MRR for different operand inputs and temperature conditions. The MRR's temperature can be increased using the integrated microheater (Fig. \ref{OSMG_gate}(a)), to consequently tune its operand-independent resonance from its fabrication-defined initial position \textit{$\gamma$} to its programmed position \textit{$\eta$}, relative to the input optical wavelength position $\lambda_{in}$ (Fig. \ref{OSMG_gate}(b)). For each bit combination at the operand terminals ((\textit{I},\textit{W}) = (0,1), (1,0), or (1,1)), the MRR's resonance passband electro-refractively moves to an operand-driven position (red and blue passbands in Fig. \ref{OSMG_gate}(b)). Based on the MRR resonance passband's programmed position \textit{$\eta$} relative to $\lambda_{in}$, the drop-port transmission (T($\lambda_{in}$)) of the MRR provides bit-wise logical AND operation between the inputs \textit{I} and \textit{W}. 

\underline{To validate our OAG}, we performed transient analysis with two pseudo-random numbers as shown in Fig. \ref{OSMG_gate}(c). For that, we modelled and simulated our OAG using the foundry-validated tools from the Ansys/Lumerical's DEVICE, CHARGE, and INTERCONNECT suites \cite{lumerical_2021}. Fig. \ref{OSMG_gate}(c) shows two input bit-streams (\textit{I}, \textit{W}) applied to the two PN junctions of our OAG at BR=10 Gbps. By looking at the output optical bit-stream T($\lambda_{in}$) in  Fig. \ref{OSMG_gate} (c), we can say T($\lambda_{in}$) = \textit{I} AND \textit{W}, which validates the functionality of our OAG as a logical AND gate. Thus, since the input bit-streams \textit{I} and \textit{W} are in the unipolar stochastic format, the output optical bit-stream at the drop port of the OAG provides the unipolar stochastic multiplication between \textit{I} and \textit{W}. 

\subsection{Photo Charge Accumulator (PCA)}\label{pcacircuit}
From Section \ref{sconna-vdpc-overview}, the stochastic multiplication bit-streams generated by OSMs are guided to a PCA, where they are accumulated to generate a binary output value equivalent to the VDP result. Our PCA is inspired from the time integrating receiver (TIR)  design from \cite{alexandermit2022} and the photodetector-based optical-pulse accumulator design from \cite{BhaskaranPCA2022}. A PCA circuit, shown in Fig \ref{SCONNA_VDPC}(c), has two stages: (i) a stochastic-to-analog conversion stage; (ii) an analog-to-binary conversion stage. The stochastic-to-analog stage employs a photodetector and two TIR circuits (one TIR circuit remains redundant, enabled by the demux and mux; Fig \ref{SCONNA_VDPC}(b)). The photodetector generates a current pulse for each optical logic `1' incident upon it. This current pulse accumulates a certain amount of charge on the capacitor of the active TIR circuit (e.g., the circuit with C1 capacitor); as a result, the capacitor accrues an analog voltage level. Hence, when one or more output optical bit-streams are incident upon the photodetector, the total accumulated charge (and thus, the accrued analog voltage level) on the active capacitor (e.g., C1) is proportional to the total number of `1's in the incident bit-streams. The number of 1's that can be accumulated in such manner might be limited, as the charge across the capacitor of TIR circuit (Fig. \ref{SCONNA_VDPC}(b)) might saturate (this is further analysed in section \ref{sec5c}). Once the TIR output saturates, a discharge of the active capacitor (e.g., C1) is needed to prepare the circuit for the next accumulation phase. While capacitor C1 is discharging, capacitor C2 of the redundant TIR circuit mitigates the discharge latency by allowing a continuation of a concurrent accumulation phase. 
The output analog voltage computed by the stochastic-to-analog conversion stage represents the unipolar unscaled addition \cite{UGEMM} of the stochastic bit-streams. To convert this analog voltage into a binary value, the analog-to-binary stage of the PCA circuit employs an analog-to-digital converter (ADC). This binary value is the VDP result.

\section{Scalability Analysis of SCONNA Architecture}\label{sectionV}
To understand the scalability of our SCONNA architecture, in this section, we analyze the achievable operating speed of the OSMs, achievable size \textit{N} of the SCONNA VDPC, and the accumulation capacity of the PCA circuits.
\subsection{Operating Speed and Latency Overhead of OSM}\label{section5a}
\begin{figure}[h!]
  \centering
  \includegraphics[width=\linewidth]{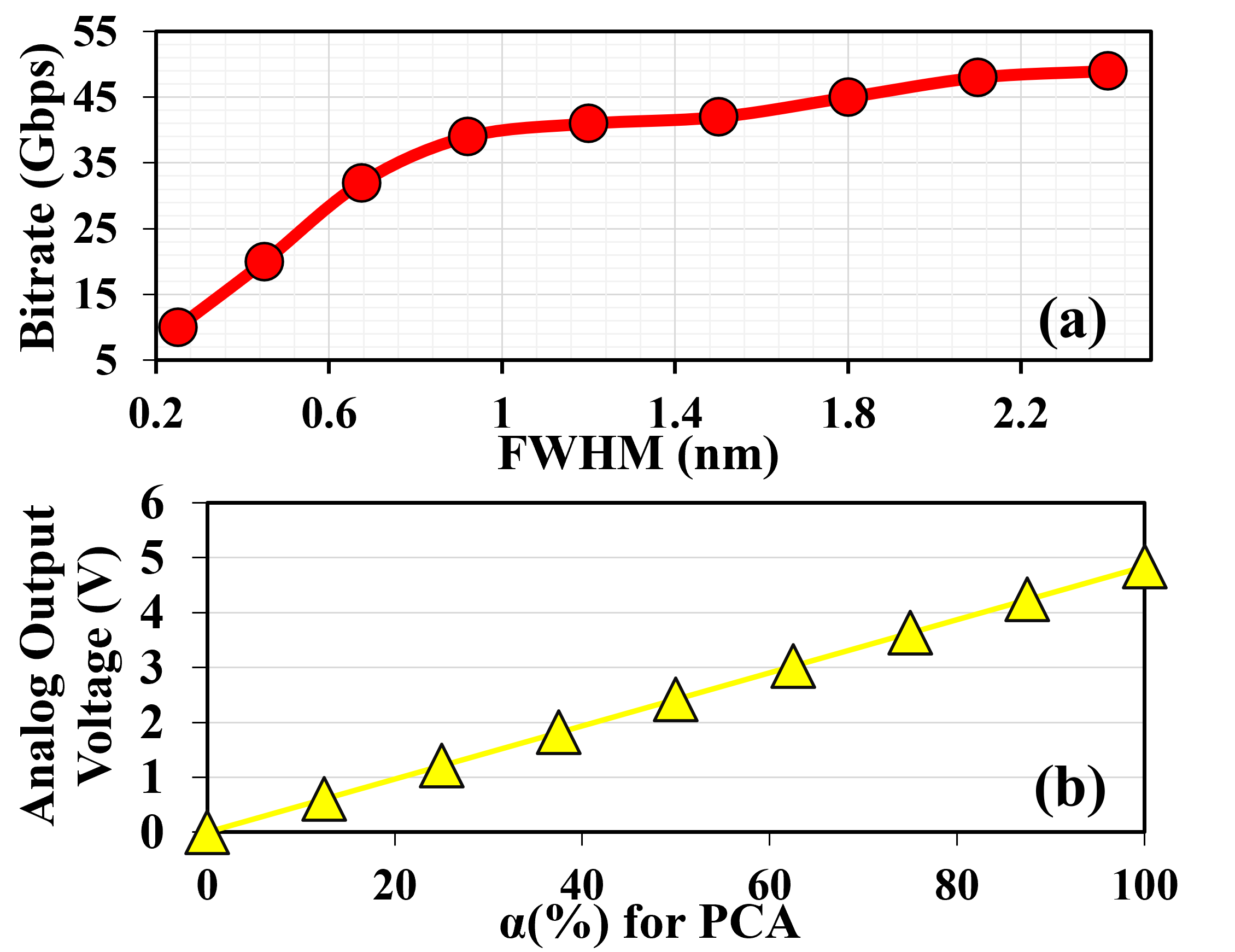}
  \caption{(a) Bitrate versus FWHM for our OSM/OAG, (b) Our PCA's analog output voltage versus $\alpha$}.   
\label{BRversusFWHM} 
\end{figure}

The peripherals of an OSM can incur some latency for accessing the scratchpad buffer and eDRAM-based lookup table. We consider 2ns latency each for the scratchpad buffer \cite{cacti7} and eDRAM-based lookup table \cite{edramlut}. Beyond this latency, the speed of an OSM depends on the achievable operating speed (bit-rate (BR)) of the constituent OAG. \underline{Analysis of OAG's BR}: For the output optical bit-stream T($\lambda_{in}$) in Fig. \ref{OSMG_gate}(c), the optical modulation amplitude (OMA) is the output power difference between the highest logic '0' power level and the lowest logic '1' power level. OMA should be at least equal to or greater than the sensitivity of the photodetector in the PCA circuit, to ensure that the photodetector in the PCA circuit can produce a distinguishably higher-amplitude current pulse for an optical logic '1' bit compared to an optical logic '0' bit. Keeping the OMA to be greater than or equal to the given photodetector sensitivity (P$_{PD-opt}$=-28dBm; Section \ref{sconnascalability}) depends on the OAG's BR and FWHM (full passband width at half maximum). Therefore, to analyze this dependency, we simulated BR and FWHM duplets for which OMA = -28 dBm, as shown in Fig. \ref{BRversusFWHM}(a). As evident, supported BR increases as FWHM increases. However, at (FWHM$\approx$0.8nm), BR saturates at 40 Gbps. Therefore, we aim to operate our OAG at BR$<=$40Gbps for FWHM$<=$0.8nm.

\subsection{Achievable Size of SCONNA VDPC}\label{sconnascalability}
We consider optimistic free-spectral range (FSR) of 50 nm \cite{lukasscalability} for the constituent MRR-based OAGs of our SCONNA VDPC. In addition, we consider the inter-wavelength gap of 0.25 nm. This allows the \textit{N} for our SCONNA VDPC to be 200 (=FSR/0.25nm), theoretically. However, even if we consider FSR=50nm to be practically achievable for our OAGs, achieving \textit{N}=200 for our SCONNA VDPC might not be possible in practice. This is because when we aim to operate our OAGs at a high BR of $<=$ 40 Gbps, for FWHM $<=$ 0.8 nm, the total power penalty for our SCONNA VDPC might increase significantly owing to the increased impacts of optical crosstalk effects at OSMs, signal truncation at MRR filters, and BR-dependent increase in the photodetector sensitivity  \cite{modulatorcrosstalk,filtersignaltruncation,pdsensitivitybr}. This increase in power penalty can reduce \textit{N} to be less than 200. Therefore, to determine the achievable \textit{N} for our SCONNA VDPC at \textit{B}=8-bit precision, we adopt the scalability analysis equations (Eq. \ref{eq2}, Eq. \ref{eq3}, and Eq. \ref{eq4}) from \cite{cases2022,lukasscalability}. Table III reports the definitions of the parameters and their
values used in these equations. Since our SCONNA VDPC processes stochastic bit-streams, which are digital in format, it requires the bit resolution of \textit{$B_{Res}$} = 1-bit in the equations. Moreover, we conservatively choose to operate OSMs/OAGs at BR=30Gbps. We consider \textit{M=N}. We first solve Eq. \ref{eq2} and Eq. \ref{eq3} for datarate (DR)=BR$*2^{B}$, to find P$_{PD-opt}$ to be -28 dBm. Then, we solve Eq. \ref{eq4} for \textit{N}, to find \textit{N}=\textit{M}=176, which is a very large \textit{N} compared to analog VDPCs that have \textit{N}$<=$44. Such large \textit{N} significantly improves the overall throughput and energy efficiency (Section \ref{evaluation}).

\begin{equation}
      B_{Res} = \frac{1}{6.02}\Bigg[20log_{10}(\frac{R\times P_{PD-opt}}{\beta\sqrt{\frac{DR}{\sqrt{2}}}}-1.76\Bigg]
      \label{eq2}
\end{equation}

\begin{equation}
    \beta = \sqrt{2q(RP_{PD-opt}+I_d)+\frac{4kT}{R_L}+R^2P_{PD-opt}^2RIN}
    \label{eq3}
\end{equation}

\begin{equation}
  \label{eq4}
\begin{split}
 P_{Laser} = \frac{10^{\frac{\eta_{WG}(dB)[N(d_{OSM})]}{10}}M}{\eta_{SMF}\eta_{EC}IL_{i/p-OSM}}
    \times\frac{P_{PD-opt}}{\eta_{WPE}IL_{MRR}}
      \\\times\frac{1}{(OBL_{OSM})^{N-1}(EL_{splitter})^{log_{2}M}} 
    \\ \times \frac{1}{(OBL_{MRR})^{N-1}(IL_{penalty})}
\end{split}
\end{equation}

\subsection{Accumulation Capacity and Error Susceptibility of PCA}\label{sec5c}
From Section \ref{sconnascalability}, our SCONNA VDPC has \textit{N}=176. For precision \textit{B}=8, each optical bit-stream in a SCONNA VDPC has $2^B$=256 bits. Therefore, each PCA in a SCONNA VDPC needs to be able to accumulate a total $N \times 2^B$=176$\times$256  optical '1' bits, at the least. We modeled the photodetector of our PCA circuit using the INTERCONNECT tool from Ansys/Lumerical \cite{lumerical_2021} for $R_{PD}$=1.2 A/W and P$_{PD-opt}$=-28 dBm, and extracted the current pulse values corresponding optical '1's and '0's that are consumed by the photodetector. We then imported these values in our MultiSim \cite{multisim} based model of the TIR circuit of the PCA, to find out that our PCA should have R=50$\Omega$, C=250pF, and voltage amplifier gain=80. For these parameters, we simulated to the analog output voltage at the PCA using MultiSim \cite{multisim} for different valus of $\alpha$=(actual \# of '1's in incident bit-streams/176$\times$256)$\times$100\%. The results are shown in Fig. \ref{BRversusFWHM}(b). As evident, the analog output voltage increases linearly with $\alpha$ without saturating at $\alpha$=100\%. This outcome shows that our PCA can efficiently support the accumulation of \textit{N}=176 bit-streams. Note that the analog output voltage from the amplifier of the PCA circuit does not incur any errors in computation. But, the ADC introduces errors in the generated binary result (we evaluate mean absolute percentage error to be 1.3\% for the ADC), and we later evaluate the impact of these errors on the CNN inference accuracy (Section \ref{evaluation}).

\begin{table*}[]
\centering
\caption{List of abbreviations and their full forms used in this paper. Definition and values of various parameters (obtained from \cite{lukasscalability}) used in Eq. \ref{eq2}, Eq. \ref{eq3}, and Eq. \ref{eq4} for the scalability analysis of our SCONNA VDPCs.}
\label{abbrevations}
\begin{tabular}{|c|c||c|c|c|}
\hline
 \textbf{Abbreviations} &  \textbf{Full   form}                                                                             &  \textbf{Parameter}             &  \textbf{Definition}                                                                                                         &  \textbf{Value}   \\ \hline
VDPC           & Vector Dot Processing Core                                                                 &  $P_{Laser}$               & Laser Power Intensity                                                                                               & 10 dBm  \\ \hline
 \textit{PCA}             &  Photo Charge Accumulator                                                         &  $R_{PD}$                     &  PD Responsivity                                                                                                     & 1.2 A/W \\ \hline
\textit{OAG}             &  Optical AND Gate                                                                     &  $R_L$                    &  Load Resistance                                                                                                     &  50 $\Omega$      \\ \hline
SE            & Summation Element                                                                                                                                         & $I_d$                    & Dark Current                                                                                                        & 35 nA   \\ \hline
SC            & Stochastic Computing                                                                  & T                    &  Absolute Temperature                                                                                                & 300 K   \\ \hline
\textit{DKV}           & Decomposed Kernel Vector                                                                & BR                    & Bitrate                                                                                                            & 30 Gbps \\ \hline
\textit{DIV}           & Decomposed Input Vector                                                                 & RIN                   & Relative Intensity Noise                                                                                            & -140 dB/Hz   \\ \hline
VDP           & Vector Dot Product                                                                      & $\eta_{WPE}$              & Wall Plug Efficiency                                                                                                & 0.1     \\ \hline
S             & Size of DKV                                                                             & $IL_{SMF}$(dB)           & Single Mode Fiber Insertion Loss                                                                                    & 0       \\ \hline
\textit{psum}          & Partial Sum                                                                           & $IL_{EC}$(dB)            & Fiber to Chip Coupling Insertion Loss                                                                               & 1.6     \\ \hline
OSM            & Optical Stochastic Multiplier                                                                 & $IL_{WG}$(dB/mm)         & Silicon Waveguide Insertion Loss                                                                                    & 0.3     \\ \hline
\textit{DR}          & Data rate                                                                             & $EL_{Splitter}$(dB)      & Splitter Insertion Loss                                                                                             & 0.01    \\ \hline
VDPE          & Vector Dot Product Element                                                              & $IL_{OSM}$(dB)           & Optical Stochastic Multiplier (OSM) Insertion Loss                                                                            &  4       \\ \hline
 \textit{N}             &  Size of VDPE                                                                            & $OBL_{OSM}$(dB)          & Out of Band Loss Optical Stochastic Multiplier                                                                                                &  0.01    \\ \hline
\textit{M}             &  Number of VDPEs per VDPC Unit                                                            &  $IL_{MRR}$(dB)           &  Microring Resonator(MRR) Insertion Loss                                                                             & 0.01  \\ \hline
OMA            & Optical Modulation Amplitude                                                                       & $IL_{penalty}$(dB) &  Network Penalty                                                                                                     &  7.3     \\ \hline
\textit{B}             & Binary Bit Precision                                                                 &  $d_{OSM}$                &  Gap between two adjacent OSMs                                                                                       & 20 $\mu$m     \\ \hline
\textit{$B_{Res}$}             & Bit Resolution                                                                 &  $P_{PD-opt}$                &  Output Photodetector Sensitivity                                                                                       & -     \\ \hline
\end{tabular}
\end{table*}

\begin{table}[]
\caption{Peripherals Parameters for Accelerators {[}6{]}.}
\label{table3}
\begin{tabular}{|cccc|}
\hline
\multicolumn{1}{|c|}{}                           & \multicolumn{1}{c|}{\textbf{Power (mW)}} & \multicolumn{1}{c|}{\textbf{Area (mm$^2$)}} & \textbf{Latency}     \\ \hline
\multicolumn{1}{|c|}{\textbf{Reduction Network}} & \multicolumn{1}{c|}{0.05}               & \multicolumn{1}{c|}{3.00E-05}           & 3.125ns              \\ \hline
\multicolumn{1}{|c|}{\textbf{Activation Unit}}   & \multicolumn{1}{c|}{0.52}               & \multicolumn{1}{c|}{6.00E-04}           & 0.78ns               \\ \hline
\multicolumn{1}{|c|}{\textbf{IO Interface}}      & \multicolumn{1}{c|}{140.18}             & \multicolumn{1}{c|}{2.44E-02}           & 0.78ns               \\ \hline
\multicolumn{1}{|c|}{\textbf{Pooling Unit}}      & \multicolumn{1}{c|}{0.4}                & \multicolumn{1}{c|}{2.40E-04}           & 3.125ns              \\ \hline
\multicolumn{1}{|c|}{\textbf{eDRAM}}             & \multicolumn{1}{c|}{41.1}               & \multicolumn{1}{c|}{1.66E-01}           & 1.56ns               \\ \hline
\multicolumn{1}{|c|}{\textbf{Bus}}               & \multicolumn{1}{c|}{7}                  & \multicolumn{1}{c|}{9.00E-03}           & 5 cycles             \\ \hline
\multicolumn{1}{|c|}{\textbf{Router}}            & \multicolumn{1}{c|}{42}                 & \multicolumn{1}{c|}{0.151}              & 2 cycles             \\ \hline
\multicolumn{4}{|c|}{\textbf{AMM/MAM}}                                                                                                                      \\ \hline
\multicolumn{1}{|c|}{\textbf{DAC \cite{dac10gbps}}}      & \multicolumn{1}{c|}{30}                 & \multicolumn{1}{c|}{0.034}              & 0.78ns               \\ \hline
\multicolumn{1}{|c|}{\textbf{ADC \cite{adc5gbps}}}      & \multicolumn{1}{c|}{29}                 & \multicolumn{1}{c|}{0.103}              & 0.78ns               \\ \hline
\multicolumn{4}{|c|}{\textbf{SCONNA}}                                                                                                                       \\ \hline
\multicolumn{1}{|c|}{\textbf{ADC \cite{adc1gbps}}}               & \multicolumn{1}{c|}{2.55}                & \multicolumn{1}{c|}{0.002}               & 0.78ns               \\ \hline
\multicolumn{1}{|c|}{\textbf{Serializer per OSM \cite{serializer}}}        & \multicolumn{1}{c|}{5}             & \multicolumn{1}{c|}{5.9}                & 0.03ns \\ \hline
\multicolumn{1}{|c|}{\textbf{LUT per OSM \cite{lut}}}    & \multicolumn{1}{c|}{0.06}            & \multicolumn{1}{c|}{0.09}               & 2ns \\ \hline
\multicolumn{1}{|c|}{\textbf{PCA \cite{multisim} }}        & \multicolumn{1}{c|}{0.02}               & \multicolumn{1}{c|}{0.28}           & -  \\ \hline
\end{tabular}
\end{table}

\section{System-Level Implementation and Evaluation}\label{evaluation}

\subsection{System-Level Implementation of SCONNA}

\begin{figure}[H]
  \centering
  \includegraphics[width=\linewidth]{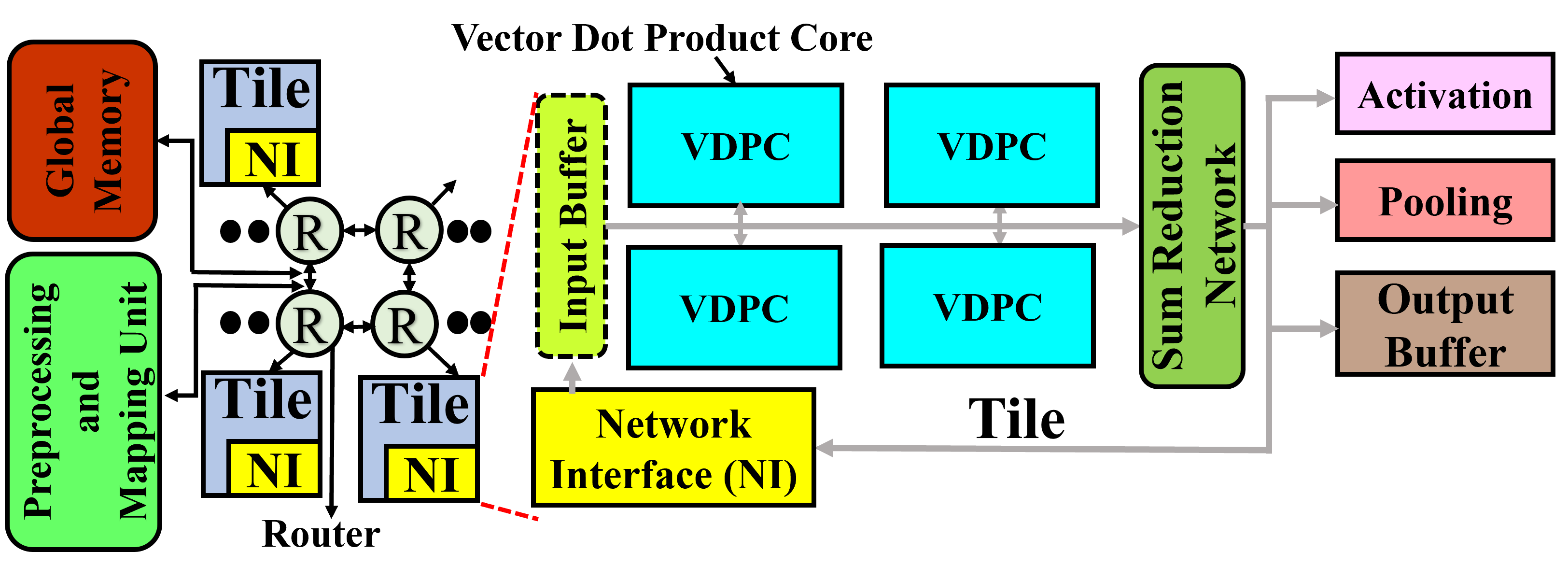}
  \caption{System-level overview of our SCONNA CNN accelerator. }   
 \label{systemlevelimplement}
\end{figure}

Fig. \ref{systemlevelimplement} illustrates the system-level implementation of our SCONNA accelerator. It consists of global memory for storing CNN parameters, and a preprocessing and mapping unit for decomposing the tensors into DIVs/DKVs and mapping them onto  VDPEs. It has a mesh of tiles connected to routers, and this mesh network facilitates parameter communication among tiles. Each tile consists of 4 SCONNA VDPCs interconnected (via H-tree network) with output buffer, activation, and pooling units. In addition, each tile also contains a \textit{psum} reduction network.

\subsection{Simulation Setup}
For evaluation, we model our SCONNA accelerator from Fig. \ref{systemlevelimplement} using our developed custom, transaction-level, event-driven  python-based simulator (\url{https://github.com/uky-UCAT/SC_ONN_SIM.git}). Using the simulator, we simulated the inference four CNN models (with batch size of 1): GoogleNet\cite{googlenet}, ResNet50\cite{resnet}, MobileNet\_V2 \cite{mobilenetv2}, and ShuffleNet\_V2 \cite{shufflenet}. We evaluate the metrics such as Frames per second (FPS), FPS/W (energy efficiency) and  FPS/W/mm$^2$ (area efficiency). We also evaluate the impact of PCA error on Top-1 and Top-5 inference accuracy of the CNN models for ImageNet validation dataset \cite{imagenet}.

We compared our accelerator with the analog optical accelerators AMM (DEAPCNN \cite{deapcnn}) and MAM (HOLYLIGHT \cite{holylight}) at 8-bits integer quantization for CNN inference. We omitted comparison with CMOS-based digital CNN accelerators as prior analog optical photonic CNN accelerators have outperformed them \cite{deapcnn,crosslight}. We simulate analog optical accelerators for 5 GS/s \cite{albireo} and from Section \ref{sec3b}, at \textit{B}=4-bit precision, we set \textit{N=16} for AMM (DEAPCNN), and \textit{N=22} for MAM (HOLYLIGHT). Prior works, AMM (DEAPCNN) and MAM (HOLYLIGHT) employ weight stationary dataflow, therefore our evaluation is based on weight stationary dataflow. For fair comparison, we perform area proportionate analysis. In the area proportionate analysis, we altered the VDPE count of each analog optical accelerator, across all of the accelerator's VDPCs, to match with the area of the SCONNA accelerator having 1024 VDPEs. The scaled VDPE count of MAM (HOLYLIGHT) and AMM (DEAPCNN) are 3971 and 3172, respectively. 

Table \ref{table3} gives the parameters used for evaluating the overhead of the peripherals in our evaluated accelerators. We consider each laser diode to emit input optical power of 10 mW (10 dBm) (Table \ref{abbrevations})\cite{deapcnn}, multiplexer and splitter parameters are taken from \cite{holylight}.

\subsection{Evaluation Results}
\begin{figure}[H]
  \centering
  \includegraphics[width=\linewidth]{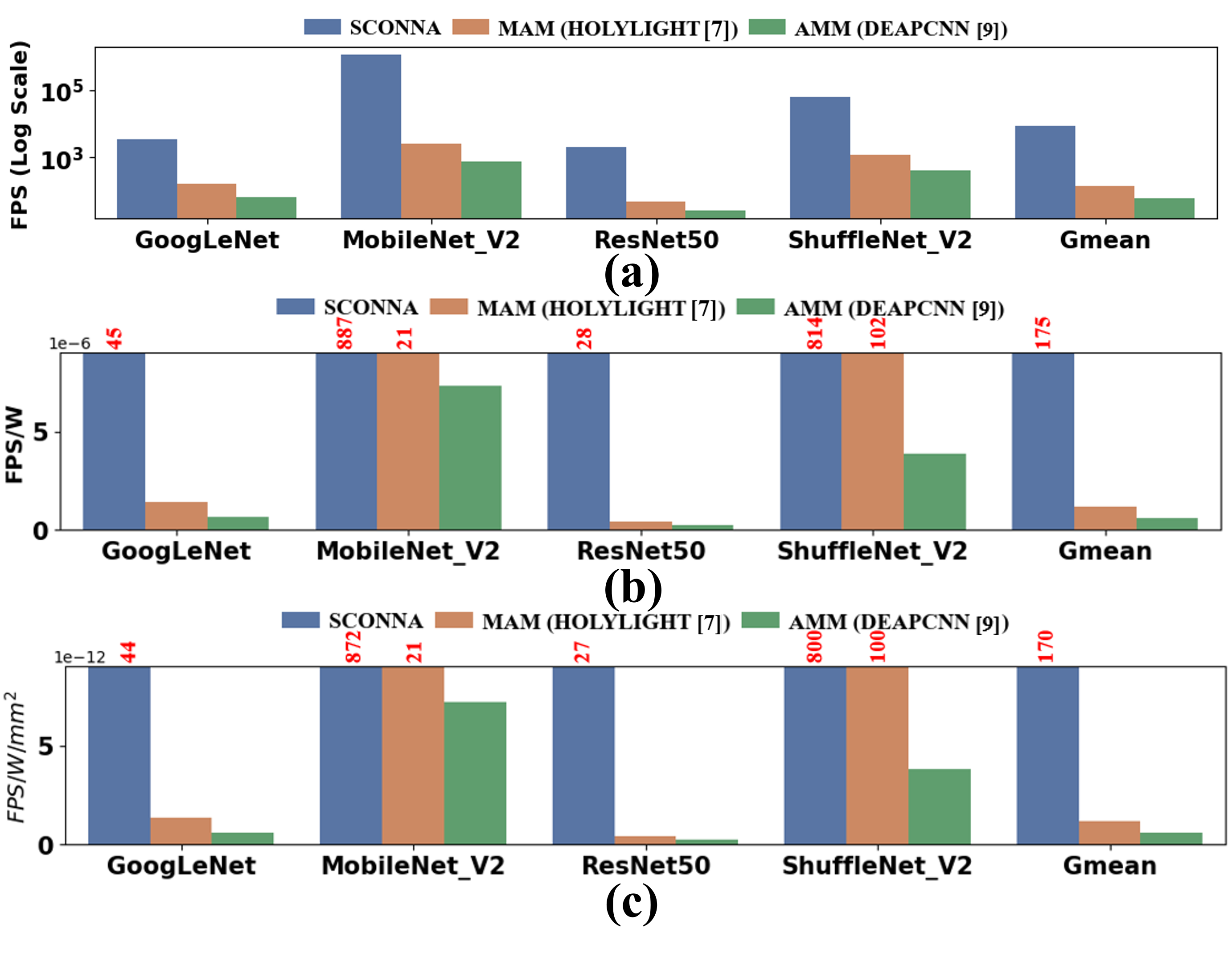}
  \caption{(a) FPS (Log Scale) (b) FPS/W (c) FPS/W/$mm^2$ for SCONNA versus MAM and AMM accelerators for \textit{B}=8-bits.}  
  \label{FPS}
\end{figure}

Fig. \ref{FPS}(a) compares the FPS values (log scale) achieved by each accelerator across various CNNs. SCONNA significantly outperforms the analog optical accelerators MAM (HOLYLIGHT) and AMM (DEAPCNN) by 66.5$\times$ and 146.4$\times$, respectively, on gmean across the CNNs. These benefits are mainly associated with the superior \textit{N} and higher BR of SCONNA compared to the analog optical accelerators. Because of the high \textit{N}, SCONNA requires less number of \textit{psums} for DKVs with \textit{S$>$44} (refer Table \ref{CNNTensorInfo}), while generating the final VDP result. The reduced \textit{psum}s drastically reduces the \textit{psum} reduction latency. The higher operating BR=30Gbps compensates for the lengthy stochastic bit-streams of $2^B$=256 bits used by SCONNA. The improvements for SCONNA are more evident for large CNNs such as GoogleNet\cite{googlenet} and ResNet50\cite{resnet} compared to smaller CNNs such as MobileNet\_V2\cite{mobilenetv2} and ShuffleNet\_V2\cite{shufflenet}. This is due to the fact that MobileNet\_V2\cite{mobilenetv2} and ShuffleNet\_V2\cite{shufflenet} employ depthwise separable convolutions which use DKVs with S$<$44 more frequently than large CNNs. Overall, SCONNA gives exceedingly better FPS compared to the analog optical accelerators.

Fig. \ref{FPS}(b) gives the energy efficiency (FPS/W) values for each accelerator across various CNNs. It is evident that SCONNA attains substantially better energy efficiency than the analog optical accelerators. Our SCONNA gains 90$\times$ and 183$\times$ better FPS/W against analog MAM (HOLYLIGHT) and AMM (DEAPCNN), respectively, on gmean across the CNNs. These energy efficiency benefits are due to the improved throughput and flexible precision support of SCONNA VDPCs. The analog MAM (HOLYLIGHT) and AMM (DEAPCNN), due to their limited 4-bit precision, employ two VDPEs to attain an 8-bit precision using bit-slicing. This decreases the throughput and also increases the energy consumption compared to SCONNA VDPCs. In addition, during area proportionate analysis, MAM (HOLYLIGHT) and AMM (DEAPCNN) get scaled to large VDPE counts (3971 and 3172), leading to overall higher static power consumption compared to SCONNA. Therefore, SCONNA achieves better energy efficiency compared to all the other tested accelerators.

Fig. \ref{FPS}(c) shows the area efficiency values (FPS/W/mm$^2$) for each accelerator across various CNNs. The area efficiency results look similar to energy efficiency as we match the area of all the accelerators to SCONNA (for the area proportionate analysis). SCONNA gains 91$\times$ and 184$\times$ better FPS/W/mm$^2$ against analog MAM (HOLYLIGHT) and AMM (DEAPCNN), respectively, on gmean across the CNNs. Overall, SCONNA significantly improves the throughput, energy efficiency and area efficiency compared to the tested analog optical accelerators.

\subsection{Inference Accuracy Results}

 As discussed in Section \ref{pcacircuit}, the ADC in the PCA circuits of our SCONNA VDPCs incurs the mean absolute percentage error of 1.3\% on the computed binary results. To evaluate the impact of these errors on the CNN inference accuracy, we simulated the inference of four CNNs on SCONNA and analog optical accelerator MAM (HOLYLIGHT). We integrated our custom simulator with ML-framework PyTorch \cite{pytorch} and performed the inference using ImageNet validation dataset \cite{2022ActiveloopHub} (50k images and 1k classes). Table \ref{accuracy} reports the Top-1 and Top-5 inference accuracies obtained for our SCONNA and MAM for four 8-bit integer-quantized CNNs. As evident, SCONNA yields Top-1 and Top-5 accuracy drop of only 0.4\% and  0.3\%, respectively, on gmean across the tested CNNs. The large CNN models ResNet50\cite{resnet} and GoogelNet\cite{googlenet} have more tolerance  to the errors, and hence, they show minimal to zero drop in accuracy for SCONNA. Furthermore, SCONNA's accuracy drop can be improved by performing stochastic computing aware training of the CNN models on SCONNA \cite{hardwareaware}. Our SCONNA accelerator's significant gains in the FPS, FPS/W, and FPS/W/mm$^2$, overshadows the minor drop in the CNN inference accuracy.

\begin{table}[]
\centering
\caption{Top-1 and Top-5 inference accuracy comparison of SCONNA versus MAM for 8-bit quantized CNNs \{GoogleNet (GNet), ResNet50 (RNet50), MobileNet\_V2 (MNet\_V2), ShuffleNet\_V2 (SNet\_V2)\} and ImageNet dataset \cite{imagenet}.}
\label{accuracy}
\begin{tabular}{|c|c|c|c|c|c|}
\hline
\begin{tabular}[c]{@{}c@{}}\textbf{SCONNA}\\\textbf{ ACCURACY} \\ \textbf{DROP} (\%)\end{tabular} &\begin{tabular}[c]{@{}c@{}}\textbf{GNet}\\ \textbf{\cite{googlenet}}\end{tabular} & \begin{tabular}[c]{@{}c@{}}\textbf{RNet}\\ \textbf{\cite{resnet}}\end{tabular} &  \begin{tabular}[c]{@{}c@{}}\textbf{MNet\_V2}\\ \textbf{\cite{mobilenetv2}}\end{tabular} & \begin{tabular}[c]{@{}c@{}}\textbf{SNet\_V2}\\ \textbf{\cite{shufflenet}}\end{tabular} & \begin{tabular}[c]{@{}c@{}}\textbf{Gmean} \end{tabular} \\ \hline
\textbf{TOP-1}                                                         & 0.1                & 0.4               & 1.5                    & 0.5                     & 0.4\\ \hline
\textbf{TOP-5}                                                         & 0.1                & 0.3               & 0.7                    & 0.4                     & 0.3\\ \hline
\end{tabular}
\end{table}

\section{Related Work on Optical CNN Accelerators}\label{relatedwork}
To accelerate CNN inferences with low latency and low energy consumption, prior works proposed various accelerators based on photonic integrated circuits (PICs) (e.g., \cite{holylight,crosslight,amm,mzi2018,mzicomplex2021}). These accelerators employ PIC-based Vector Dot Product Cores (VDPCs) to perform multiple parallel VDP operations. Some accelerators implement digital VDPCs (e.g., \cite{pixel,albireo}), whereas some others employ analog VDPCs (e.g., \cite{holylight,crosslight,deapcnn,tait2017}). Different VDPC implementations employ MRRs (e.g., \cite{holylight,deapcnn,crosslight,tait2020photonic,feldmannMRR2021parallel}) or MZIs (e.g., \cite{cansu2021,mzi2018,mzicomplex2021}) or both (e.g., \cite{pixel}, \cite{albireo}). The analog VDPCs can be further classified as incoherent (e.g., \cite{holylight,crosslight,deapcnn}) or coherent (e.g., \cite{coherentrayhamerly2019,coherentZhao2019,coherentnature21,coherentZhao18,coherent2018,chorentzhou2021large}). To set and update the values of the individual input and kernel tensors used for vector dot product operations, the incoherent VDPCs utilize the analog optical signal power, whereas the coherent VDPCs utilize the electrical field amplitude and phase. The coherent VDPCs achieve low inference latency, but they suffer from control complexity, high area overhead, low scalability, low flexibility, high encoding noise, and phase error accumulation issues \cite{limitsCohorent}. In contrast, the MRRs-enabled incoherent VDPCs based accelerators achieve better scalability and lower footprint, because they use PICs that are based on compact MRRs \cite{deapcnn}, unlike the coherent VDPCs that use PICs based on bulky MZIs. Various state-of-the-art PIC-based optical CNN accelerators are well discussed in a survey paper \cite{acceleratorssurvey}. Because of the inherent advantages of MRR-enabled incoherent VDPCs, there is impetus to design more energy-efficient and scalable CNN accelerators employing MRR-enabled incohorent VDPCs.

\section{Conclusions}
 To mitigate the very strong scalability versus bit-precision trade-off innately present in analog optical CNN accelerators, we demonstrated a merger of stochastic computing and MRR-based CNN accelerators for the first time in this paper. We invented an MRR-based optical stochastic multiplier (OSM) and employed multiple OSMs to forge a novel stochastic computing based CNN accelerator called SCONNA. Our evaluation results for four CNN models show that SCONNA provides improvements of up to 66.5$\times$, 90$\times$, and 91$\times$ in throughput, energy efficiency, and area efficiency, respectively, compared to two analog optical accelerators AMM and MAM, with Top-1 accuracy drop of only up to 0.4\% for large CNNs and up to 1.5\% for small CNNs.

\section*{Acknowledgments}
We thank the anonymous reviewers whose valuable feedback helped us improve this paper. We would also like to acknowledge the National Science Foundation (NSF) as this research was supported by NSF under grant CNS-2139167.

\bibliographystyle{IEEEtran}
\bibliography{references}

\vfill

\end{document}